# Growth of superconducting $MgB_2$ thin films via postannealing techniques

W.N. Kang[*], Eun-Mi Choi, Hyeong-Jin Kim, Hyun-Jung Kim, and Sung-Ik Lee

*National Creative Research Initiative Center for Superconductivity, Department of Physics, Pohang University of Science and Technology, Kyungbuk 790-784, South Korea*

We report the effect of annealing on the superconductivity of $MgB_2$ thin films as functions of the postannealing temperature in the range from 700°C to 950°C and of the postannealing time in the range from 30 min to 120 min. On annealing at 900°C for 30 min, we obtained the best-quality $MgB_2$ films with a transition temperature of 39 K and a critical current density of ~ $10^7$ A/cm$^2$. Using the scanning electron microscopy, we also investigated the film growth mechanism. The samples annealed at higher temperatures showed the larger grain sizes, well-aligned crystal structures with preferential orientations along the c-axis, and smooth surface morphologies. However, a longer annealing time prevented the alignment of grains and reduced the superconductivity, indicating a strong interfacial reaction between the substrate and the $MgB_2$ film.

*Corresponding author. Tel: +82-54-279-5847; Fax: +82-54-279-5299.
E-mail address: wnkang@postech.ac.kr (WNK) or silee@postech.ac.kr (SIL)

## 1. Introduction

The discovery of superconductivity in the $MgB_2$ compound [1], with the highest transition temperature ($T_c$) of 39 K among the metallic superconductors, has generated great interest in both basic scientific [2-9] and practical applications [10-17]. Recently, the two-gap nature of the $MgB_2$ superconductor has been confirmed by using scanning tunneling microscopy on highly c-axis oriented thin films. The strongly linked nature of the inter grains [11] with metallic transport properties [4-9] and the relatively simple crystal structure in this material are further advantages of its use in technological applications. An upper critical field ($H_{c2}$) of 29 - 39 T [4,10], which is much higher than those of conventional superconductors, was observed, suggesting that $MgB_2$ would be very promising compound for large-scale application in superconducting solenoids using mechanical cryocoolers, such as closed cycle refrigerators. In addition to its higher $T_c$ and $H_{c2}$, the magnitude of the critical current density ($J_c$) is also a very important factor for use in practical applications. In the case of $MgB_2$ thin films with preferred orientations along the c-axis, a very high-current carrying capability, $J_c$ ~ $10^7$ A/cm$^2$ at 5 K, was observed [9], which is two orders of magnitude larger than the values for the Fe-clad $MgB_2$ wires prepared by using powder in tube techniques [12-15].

Since the successful growth of $MgB_2$ thin films [14], a number of groups have intensively conducted experiments to fabricate high-quality films [17-28]. Two-step (*ex-situ*) techniques, which require Mg diffusion into pure B films at high temperatures, have been quite successful in growing high-quality films [15-18]. However, the surface quality for electronic-device applications is still far from being achieved. Nevertheless, very remarkable research results concerning electronic-device applications, such as superconducting quantum interference devices (SQUID) and microwave devices, have been reported [28-31]. Brinkman *et al.* [28] reported successful fabrication of $MgB_2$ SQUID. Using a low-$T_c$ thin films ($T_c$ ~ 22 K), they obtained a nice voltage modulation up to 19 K, showing that high-temperature SQUID operating above 30 K could be developed by using metallic superconductors if higher-$T_c$ MgB$_2$ films are used. Moreover, Klein *et al.* [30] observed a very low surface resistance of 19 μΩ cm at 7.2 GHz and 4.2 K in MgB$_2$ films, which is comparable with that of a high-quality YBa$_2$Cu$_3$O$_7$ thin film. This is a very surprising result since MgB$_2$ films have a rough surface morphology compared to YBa$_2$Cu$_3$O$_7$ thin films. Very recently, *in-situ* growth of high-quality MgB$_2$ thin films by using a hybrid physical-chemical vapor deposition (HPCVD) technique, has been reported by Zeng *et al.* [32]. This is believed to be a very promising growth technique in terms of widespread applications for superconducting electronics, such as multilayer Josephson junctions and digital circuits.

In this paper, we report the superconducting properties and the growth mechanism of MgB$_2$ thin films prepared under various annealing temperatures and times. The films annealed at 900°C for 30 min showed the best superconductivity whereas the samples annealed at 950°C for 30 min revealed smooth and compact surface morphologies with preferential orientations along the c-axis. The present study was possible since we used a pulsed laser deposition (PLD) system attached a load-lock sample preparation chamber, which routinely fabricated high-quality MgB$_2$ thin films with reliable reproducibility.

**2. Experiment**

The MgB$_2$ thin films were grown on r-cut Al$_2$O$_3$ (1 $\bar{1}$ 0 2) single crystals in a high-vacuum condition at ~ 10$^{-7}$ Torr by using the PLD and the postannealing techniques, reported in our earlier paper [9,16]. The PLD system used in the present study is especially designed for the growth of metallic thin films, which should not be contaminated by O$_2$ or H$_2$O. Furthermore, this system is able to transfer the precursor B films from the growth chamber to a clean dry box without exposure to air. We found that the preparation of a pure B film was very crucial for fabricating of high-quality c-axis-oriented MgB$_2$ thin films. The laser energy density used in this study was 10 – 20 J/cm$^2$ at a laser flux of 450 mJ/pulse and a pulse frequency of 8 Hz. After precursor B films had been deposited, it was put into an Nb tube together with high-purity Mg metal (99.9%), and the tube was sealed by using arc welder in an Ar atmosphere. The postannealing was carried out at several temperatures from 700°C to 950°C and at several annealing times from 30 to 120 min The film growth conditions and transport properties are summarized in Table I. Typical dimensions of the samples were 10 mm in length, 10 mm in width and 0.4 – 0.5 μm in thickness.

The resistivity was measured by standard dc four-probe method after cutting the samples into rectangular shapes with a size of 1 mm x 5 mm. To obtain good ohmic contacts, we coated Au film on a contact pad after cleaning the film surface by using Ar-ion beam milling. This process doesn't reduce the superconducting properties of MgB$_2$ thin films. The magnetization was measured by using magnetometry. The surface morphology was studied by using field-emission scanning electron microscopy (SEM).

**3. Results and discussion**

Figures 1(a) – (d) show SEM pictures for various MgB$_2$ films annealed for 30 min at (a) 800°C, (b) 850°C, (c) 900°C, and (d) 950°C, and annealed at 900°C for (e) 60 min and (f) 120 min. Sub-micron sized MgB$_2$ single crystals having hexagonal shape can be observed. We can see clearly that the grain connectivity becomes very strong as the annealing temperature is increased. The higher temperature also assists to form preferred orientation along the c-axis, which is perpendicular to substrate surface, whereas annealing times longer than 60 min at 900°C hinders the alignment of grains so that a polycrystalline surface morphology is revealed, as shown in Figs. 1(e) and (f). It is very interesting that MgB$_2$ thin films grow with preferential orientations on Al$_2$O$_3$ even though the lattice-matching relationship between the MgB$_2$ and the substrate is not well satisfied. Recently, Tian *et al.* [25] reported an interesting result based on interfacial reaction between the MgB$_2$ film and the Al$_2$O$_3$ substrate. They found intermediate epitaxial layers of MgO and MgAl$_2$O$_4$ in the film/substrate

interface so that epitaxial growth of $MgB_2$ on an $Al_2O_3$ substrate is possible. Our results further suggest that $MgB_2$ thin films grow along a favorable direction in a wide temperature window of 850 – 950°C and for an annealing time around 30 min. If the polycrystalline growth at longer annealing times shown in Figs. 1(e) and (f) is to be understood, further microscopic experiments addressing the crystallographic relations of $MgB_2$ on $Al_2O_3$ substrates is required.

The temperature dependence of the resistivity of the $MgB_2$ films after annealing (a) at 700°C – 950°C for 30 min and (b) at 900°C for 30 – 120 min, are shown in Fig. 2, where the resistivity is plotted using values normalized to the value at 290 K. The transport properties extracted from these data are listed in Table I. All samples show higher values for $T_c$, 37.4 – 39.1 K, whereas the transition width ($\Delta T_c$) varies significantly with annealing conditions. Samples with smaller $\Delta T_c$ and $\rho_{290K}$ show better superconductivity; thus, the annealing at 900°C for 30 min can be considered as an optimum growth process. As the annealing temperature is increased up to 900°C, we can see a gradual increase in the RRR and a dramatic decrease in the normal-state resistivity, indicating that the grain connectivity becomes very strong with increasing temperature, as indicated in the SEM images in Fig. 1. However, M950C30m has a much higher resistivity than M900C30m although the SEM image shows strong connectivity of the grain boundary. This suggests that an interfacial reaction between the substrate and the $MgB_2$ film can result from high-temperature annealing at temperatures above 900°C. For the sample annealed at 900°C, the resistivity increases gradually with increasing annealing time from 30 min to 120 min, which is consistent with the polycrystalline surface morphology shown in Figs. 1(e) and (f).

Figure 3 shows the temperature dependence of the zero-field-cooled magnetization data at 10 Oe for the samples fabricated (a) at 700°C – 950°C for 30 min and (b) at 900°C for 30 – 120 min Regardless of the growth condition, all samples show bulk diamagnetism at temperatures below 10 K. For the samples annealed for 30 min, we can see a systematic enhancement of $T_c$ with increasing annealing temperature up to 900°C. The $T_c$ of M950C30m is lower than that of M900C30m, but is comparable to that of single crystalline $MgB_2$. Considering the sharp $\Delta T_c$ of ~ 0.3 K for M950C30m and its well-grown surface morphology, as shown in Fig. 1(d), M950C30m may be a good sample for studying physical properties. However, M900C30m better satisfies the requirements for practical applications because of its higher $T_c$ and $J_c$ (see Fig. 4). Using M950C30m, indeed, decisive experimental result was obtained by Iavarone *et al.* [2], confirming the two-gap nature of the $MgB_2$ superconductor. In Fig. 3(b), the samples annealed for 60 – 120 min show a lower $T_c$ and a broad superconducting transition, indicating that longer annealing degrades superconductivity by changing the growth orientation of $MgB_2$ grains, as shown in the SEM images.

Figure 4 shows the magnetic field dependence of $J_c$ at 5 K for M700C30m, M800C30m, M900C30m, M950C30m, M900C60m, and M900C120m. To estimate the critical current density, we measured the M–H loop. The sample size, $5 \times 3 - 4$ mm$^2$, rather than the grain size, was used to evaluate $J_c$ by using Bean's model. M900C30m had the highest $J_c$ of all the samples, $2.5 \times 10^7$ A/cm$^2$ at 5 K in zero field, which is comparable to the values in previous reports on $MgB_2$ films [7,19,32] and on other high-$T_c$ cuprates [33,34]. A $J_c$ of ~ $10^6$ A/cm$^2$ at 4.5 T is sufficiently high for practical applications to high-field superconducting magnet systems. This result reflects that the grains of M900C30m are connected very strongly with a high density of pinning sites. At zero field, the $J_c$ values for the other films are ~ $6 \times 10^6$ A/cm$^2$ at zero field, but the magnetic field dependence of $J_c$ shows quite different behavior. The samples annealed at equal or below 900°C and for shorter time (30 min), show weak-field dependence compared to the samples of M950C30m and M900C120m. As shown in the SEM image [Fig. 1(f)], M900C120m has a polycrystalline structure; thus, the $J_c$ depends strongly on the magnetic field as with polycrystalline Fe-clad $MgB_2$ wires [13]. Different from M900C120m, however, M950C30m shows a very dense surface morphology highly oriented along the c-axis as shown in Fig. 1(d). In the case of this film, the strong-field dependence of $J_c$ probably implies that this sample contains fewer pinning sites than M900C30m [Fig. 1(c)]. Our $J_c$ data further support the fabrication process for M900C30m sample being the optimal condition for large-scale applications. For electronic-device application, M950C30m is favored, but we should reduce a possible interfacial reaction at high temperatures. A possible solution for this difficulty is on *in-situ* low-temperature process [26,27,32] using chemically stable substrates or buffer layers.

## 4. Summary

The superconducting properties and the growth mechanism of $MgB_2$ thin films prepared under various annealing temperatures and times were investigated. The films annealed at 900°C for 30 min showed the best superconductivity with a transition temperature of 39 K and a critical current density of ~ $10^7$ A/cm$^2$ at 5 K and zero field. On the other hand, the samples annealed at 950°C for 30 min revealed smooth and compact surface morphologies with preferential orientations along the c-axis. We found that $MgB_2$ thin films grow along a preferred direction in the annealing temperature window of 850 – 950°C and for annealing times around 30 min. By using a PLD system attached a load-lock sample preparation chamber, we are able to fabricate very high-quality $MgB_2$ thin films with reliable reproducibility. Our results suggest that an *in-situ* low-temperature process should be developed in order to apply $MgB_2$ thin films to superconducting electronics.

Acknowledgement

This work is supported by the Creative Research Initiatives of the Korean Ministry of Science and Technology.

**Figure captions**

Fig. 1. SEM pictures for the $MgB_2$ films annealed for 30 min at (a) 800°C, (b) 850°C, (c) 900°C, and (d) 950°C, and annealed at 900°C for (e) 60 min and (f) 120 min.

Fig. 2. Temperature dependence of the resistivity for $MgB_2$ films after annealing (a) at 700°C – 950°C for 30 min and (b) at 900°C for 30 – 120 min.

Fig. 3. Temperature dependence of zero-field-cooled magnetization data at 10 Oe for the films fabricated (a) at 700°C – 950°C for 30 min and (b) at 900°C for 30 – 120 min.

Fig. 4. Magnetic field dependence of $J_c$ at 5 K for various films grown at 700°C – 950°C for 30 min and at 900°C for 30 – 120 min.

Table I. Summary of the film growth conditions, annealing temperature ($T_{ann}$) and annealing time ($t_{ann}$), and the transport properties superconducting transition temperature ($T_c$), transition width ($\Delta T_c$), normal-state resistivity ($\rho_{290K}$) at 290 K, and residual resistivity ratio (RRR = $\rho_{40K}/\rho_{290K}$).

| Sample ID | $T_{ann}$ (K) | $t_{ann}$ (min) | $T_c$ (K) | $\Delta T_c$ (K) | $\rho_{290K}$ ($\mu\Omega$ cm) | RRR |
| --- | --- | --- | --- | --- | --- | --- |
| M700C30m | 700 | 30 | 38.1 | 2.3 | 68 | 1.4 |
| M800C30m | 800 | 30 | 37.5 | 0.4 | 75 | 1.5 |
| M850C30m | 850 | 30 | 37.4 | 0.7 | 49 | 1.7 |
| M900C30m | 900 | 30 | 39.0 | 0.3 | 14 | 2.3 |
| M950C30m | 950 | 30 | 37.5 | 0.3 | 24 | 2.1 |
| M900C60m | 900 | 60 | 39.1 | 0.6 | 18 | 2.9 |
| M900C120m | 900 | 120 | 38.3 | 1.5 | 20 | 2.1 |

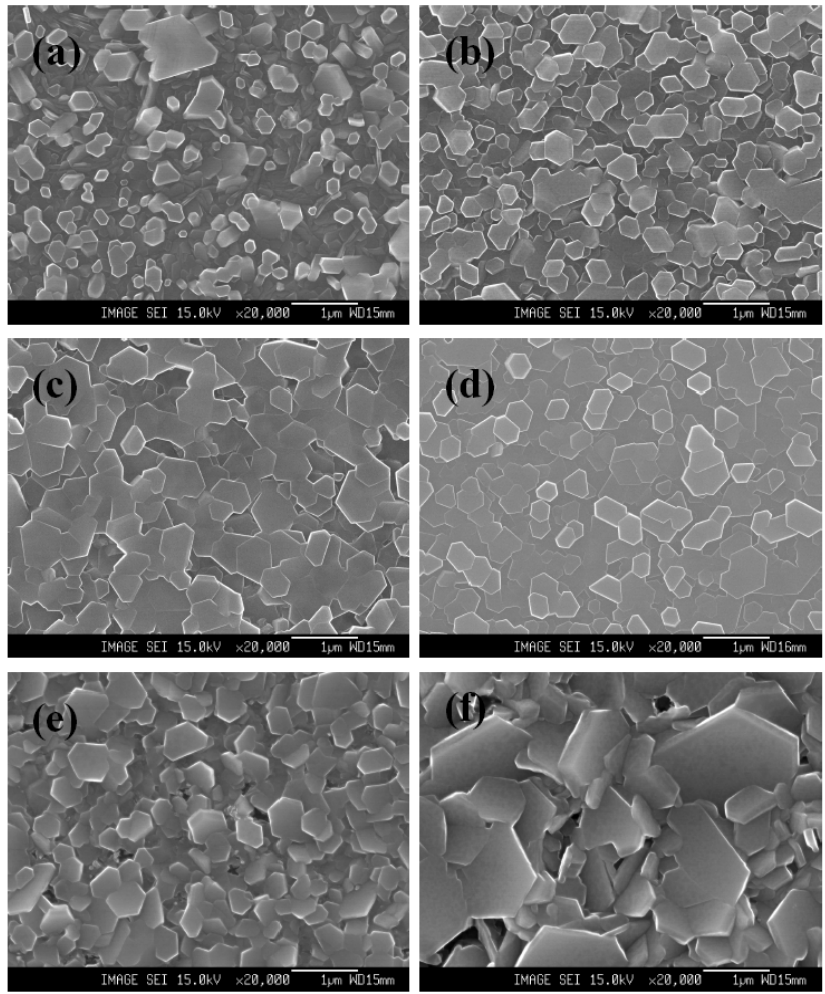

Fig. 1. Kang et al.

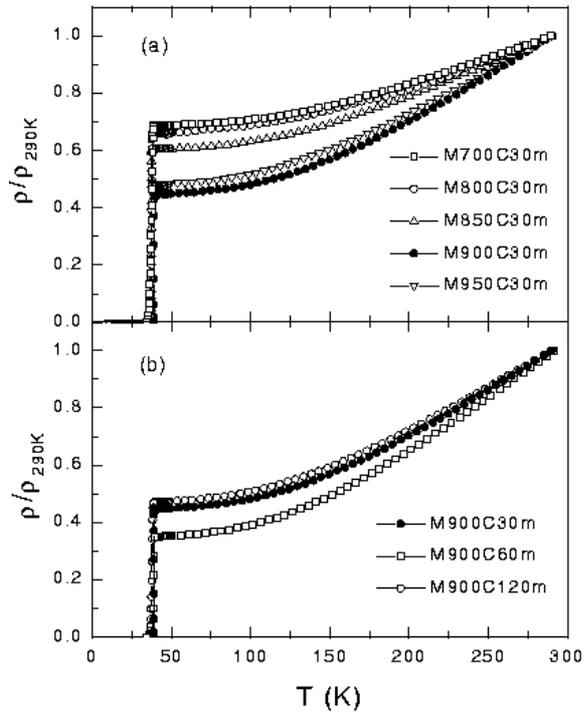

Fig. 2. Kang et al.

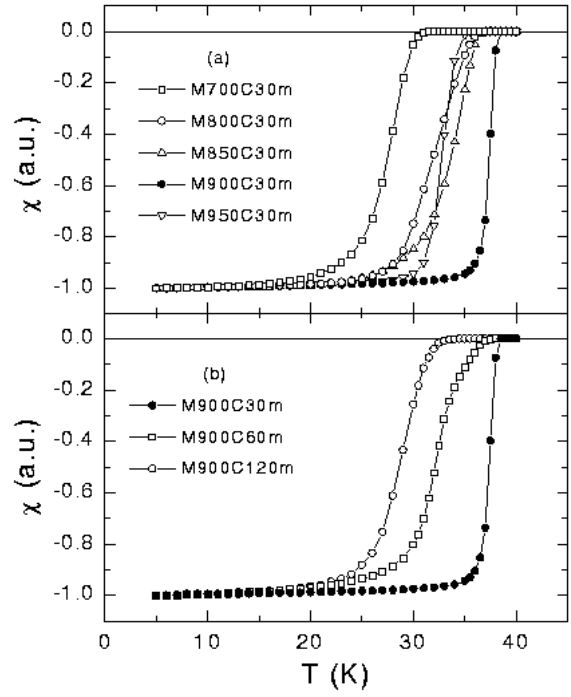

Fig. 3. Kang et al.

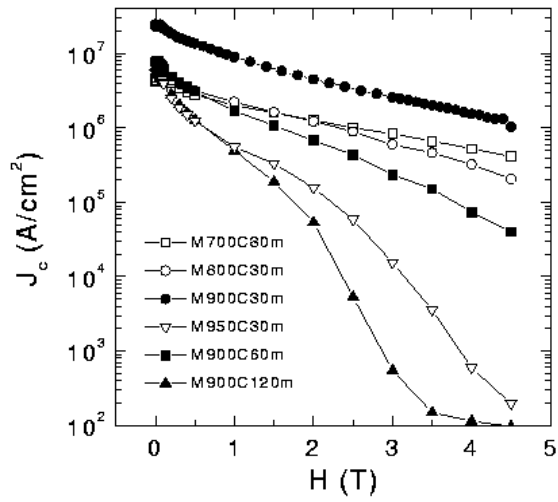

Fig. 4. Kang et al.

7